\documentclass[letterpaper]{article} 
\usepackage{aaai2026}  
\usepackage{times}  
\usepackage{helvet}  
\usepackage{courier}  
\usepackage[hyphens]{url}  
\usepackage{graphicx} 
\usepackage{natbib}  
\usepackage{caption} 
\usepackage{algorithm}
\usepackage{algorithmic}

\usepackage{newfloat}
\usepackage{listings}
\DeclareCaptionStyle{ruled}{labelfont=normalfont,labelsep=colon,strut=off} 
\floatstyle{ruled}
\newfloat{listing}{tb}{lst}{}
\floatname{listing}{Listing}
\usepackage{multirow}
\usepackage{tabularx}
\usepackage{booktabs}
\usepackage{amsfonts}
\usepackage{amsmath}
\usepackage{adjustbox}

\title{HQ-SVC: Towards High-Quality Zero-Shot Singing Voice Conversion in Low-Resource Scenarios}
\author{
    Bingsong Bai\textsuperscript{\rm},
    Yizhong Geng\textsuperscript{\rm},
    Fengping Wang\textsuperscript{\rm},
    Cong Wang\textsuperscript{\rm},
    Puyuan Guo\textsuperscript{\rm},\\
    Yingming Gao\textsuperscript{\rm},
    Ya Li\textsuperscript{\rm}\thanks{Corresponding author.}
}
\affiliations{
    \textsuperscript{\rm}Beijing University of Posts and Telecommunications, China\\
    \{bingsongbai, yzgeng, wfp, congwang, guopy, yingming.gao, yli01\}@bupt.edu.cn
}

\usepackage{bibentry}

\begin{document}

\maketitle

\begin{abstract}
Zero-shot singing voice conversion (SVC) transforms a source singer's timbre to an unseen target speaker's voice while preserving melodic content without fine-tuning. Existing methods model speaker timbre and vocal content separately, losing essential acoustic information that degrades output quality while requiring significant computational resources. To overcome these limitations, we propose HQ-SVC, an efficient framework for high-quality zero-shot SVC. HQ-SVC first extracts jointly content and speaker features using a decoupled codec. It then enhances fidelity through pitch and volume modeling, preserving critical acoustic information typically lost in separate modeling approaches, and progressively refines outputs via differentiable signal processing and diffusion techniques. Evaluations confirm HQ-SVC significantly outperforms state-of-the-art zero-shot SVC methods in conversion quality and efficiency. Beyond voice conversion, HQ-SVC achieves superior voice naturalness compared to specialized audio super-resolution methods while natively supporting voice super-resolution tasks.
\end{abstract}

\begin{links}
    \link{Code}{https://github.com/ShawnPi233/HQ-SVC}
\end{links}

\section{Introduction}
\label{sec:intro}
Zero-shot singing voice conversion (SVC) aims to transform a singer's timbre to that of another unseen target speaker while preserving the melody and content, without requiring prior data or fine-tuning for the target speaker. It can be widely applied in music production and virtual singers. However, zero-shot singing voice conversion faces challenges due to the high sampling rates and broader fundamental frequency (F0) ranges in singing voices, making singing voice synthesis more difficult. Additionally, the intricate coupling between timbre and pitch further complicates the task. Moreover, the limited availability of high-quality singing datasets and computational resources highlights the need of studying zero-shot singing voice conversion in low-resource scenarios.

Existing SVC methods \cite{liu2021diffsvc,so-vits-svc2023,zhou2023vitssvcc,bai24_spa_svc} have demonstrated high-quality singing voice conversion by leveraging generative models like VITS \cite{kim2021conditional_vits} and diffusion \cite{ho2020ddpm} models. However, these models typically rely on explicit speaker IDs and separate content encoders \cite{hsu2021hubert,qian2022contentvec,chen2022wavlm}, which limit their generalization ability to unseen speakers, making them unsuitable for zero-shot singing voice conversion. FastSVC \cite{liu2021fastsvc_cross} achieves zero-shot singing voice conversion with a lightweight model but struggles with poor synthesis quality. LDM-SVC \cite{chen2024ldm} provides high-quality zero-shot singing voice conversion on small datasets but demands significant computational resources and training time due to its two-stage and adversarial training. SaMoye-SVC \cite{wang2024samoye} also delivers high-quality zero-shot singing voice conversion but requires large datasets and extensive adversarial training, making it highly resource-intensive. Moreover, these methods often rely on separate content and speaker encoders, leading to information loss and hindering effective feature fusion and reconstruction of natural and expressive voice.

Given that both speech and singing belong to voice, the inherent similarities between them allow large-scale speech data to provide valuable priors, enabling transfer learning methods to effectively improve performance in zero-shot singing voice conversion tasks. Recent advancements in vector quantization (VQ) for zero-shot text-to-speech models like NaturalSpeech3 \cite{junaturalspeech3} and UniCATS \cite{du2024unicats} have achieved significant success. NaturalSpeech3 introduces a VQ-based audio codec, FACodec, which turn audio into distinct latent spaces for content and speaker features simultaneously. This architecture allows zero-shot disentanglement of speaker and content features for unseen speakers, while also reducing information loss and making feature alignment easier compared to separate modeling. Although FACodec enables effective disentanglement of speaker and content features, it does not fully capture the complex acoustic variations necessary for high-quality synthesis, especially in challenging zero-shot singing voice conversion tasks. 

Therefore, based on unified decoupled audio codec, we propose HQ-SVC, a novel framework designed to achieve high-quality zero-shot singing voice conversion in low-resource scenarios. Additionally, to improve high-quality synthesis modeling, we first leverage decoupled codec’s zero-shot disentanglement capability to extract content and speaker features from the decoder’s intermediate layer outputs. We then introduce supplementary features to  improve the performance of SVC. To further enhance feature disentanglement and fusion, we propose the Enhanced Voice Adaptation (EVA) module, which incorporates additional acoustic features for multi-feature fusion, introduces a speaker loss, and includes a Speaker-F0 Predictor to enhance synthesis quality. Following the approach in DDSP-SVC \cite{ddsp-svc2023}, the synthesis process is progressively optimized using Differentiable Digital Signal Processing (DDSP) \cite{engel2020ddsp} and diffusion \cite{ho2020ddpm} models to generate Mel spectrograms. Finally, a vocoder is employed to produce singing voices. 

By avoiding unnecessary retraining and adversarial training, HQ-SVC is lightweight and time-efficient, enabling zero-shot singing voice conversion for unseen speakers with low computational resources and small datasets. Additionally, since the input features are extracted at a low sampling rate and aligned with high-sampling-rate Mel spectrograms, HQ-SVC also supports zero-shot voice super-resolution (SR). Therefore, the main contributions of this work are summarized as follows:

\begin{itemize}
\item We propose HQ-SVC, which achieves high-quality zero-shot singing voice conversion under low-resource scenarios, such as training on a single consumer-grade GPU and small-scale datasets. HQ-SVC outperforms the current SOTA method, across most objective and subjective evaluation metrics.
\item We introduce the EVA module to improve multi-feature fusion and enhance synthesis quality. It incorporates speaker loss and a Speaker-F0 Predictor, enabling more natural singing voice conversion.  
\item HQ-SVC extends its capabilities to zero-shot singing and speech SR tasks. Compared to the current SOTA audio SR method, HQ-SVC generates voice with higher naturalness and speaker similarity. Our work may inspire further research on using zero-shot singing voice conversion methods to achieve higher-quality voice SR tasks.
\end{itemize}

\section{Related Works}
\subsection{Singing Voice Conversion}
Singing Voice Conversion (SVC) aims to transform a singer’s timbre into that of another target singer while preserving the melody and content. Early SVC methods \cite{doi2012gmmsvc,kobayashi2014gmmsvc,kobayashi2015gmmsvc} were primarily based on Gaussian Mixture Models (GMM), which relied on parallel data. These methods require both source and target singers to perform the same song. This dependency not only demanded a large amount of parallel data but also limited the generation quality of the models. With the rise of Generative Adversarial Networks (GANs) \cite{goodfellow2020gan}, several GAN-based SVC approaches \cite{sisman2019singan,sisman2020generative,liu2021fastsvc_cross} emerged that operate without parallel data. These methods employ adversarial training, where a generator and a discriminator engage in a game to improve the audio quality. However, GAN-based methods typically take longer to train and can suffer from issues like mode collapse, making the overall training process more challenging. SoVITS-SVC \cite{so-vits-svc2023} builds on adversarial training by adding a flow-based model to prevent mode collapse during training. It is based on VITS \cite{kim2021conditional_vits}, which uses an end-to-end flow-based architecture with adversarial training to generate high-quality singing voices. While effective, it still requires considerable computational resources for training.

\subsection{High-Quality SVC in Low-resource Scenarios}
Open-source singing voice datasets available for academic research remain relatively scarce. At the same time, the growing practical demand for efficient training and inference on consumer-grade GPUs has led to increasing research attention on achieving high-quality SVC under low-resource conditions. Low-resource SVC means using only a small amount of singing voice data and limited GPU power during training, while still producing high-quality audio.
Several methods have been proposed to address this challenge. These include DDSP-SVC \cite{ddsp-svc2023} based on Differentiable Digital Signal Processing (DDSP) models \cite{engel2020ddsp} and diffusion-based SVC methods\cite{liu2021diffsvc,lu2024comosvc,huang2025lhq}. These methods not only demonstrate superior generation quality, but also exhibit strong training stability, enabling high-quality singing voice conversion even in low-resource scenarios.

\subsection{Zero-shot SVC}
Nevertheless, while the aforementioned SVC methods have made significant strides, their reliance on explicit speaker identities restricts their ability to generalize to unseen speakers, making them unsuitable for zero-shot SVC scenarios. In zero-shot SVC, the objective is to achieve high-quality conversion without requiring target singer data or fine-tuning. Traditional SVC approaches typically depend on parallel data and explicit speaker labels, and in practical applications, issues such as high sampling rates, broader pitch ranges, and the intricate coupling between timbre and pitch present significant challenges in modeling and training. FastSVC \cite{liu2021fastsvc_cross} is one of the earlier zero-shot singing voice conversion methods; it leverages GANs combined with FiLM \cite{perez2018film} to enable a lightweight zero-shot conversion. However, its generation quality leaves room for improvement. In contrast, methods like LDM-SVC \cite{chen2024ldm}, FreeSVC \cite{ferreira2025freesvc}, and SaMoye-SVC \cite{wang2024samoye}, which are based on VITS, can synthesize high-quality singing voices in a zero-shot manner, although at the cost of requiring considerable computational power during training.

\section{Methods}
\label{sec:hqsvc}

\begin{figure*}[t]
    \centering
        \includegraphics[width=\linewidth]{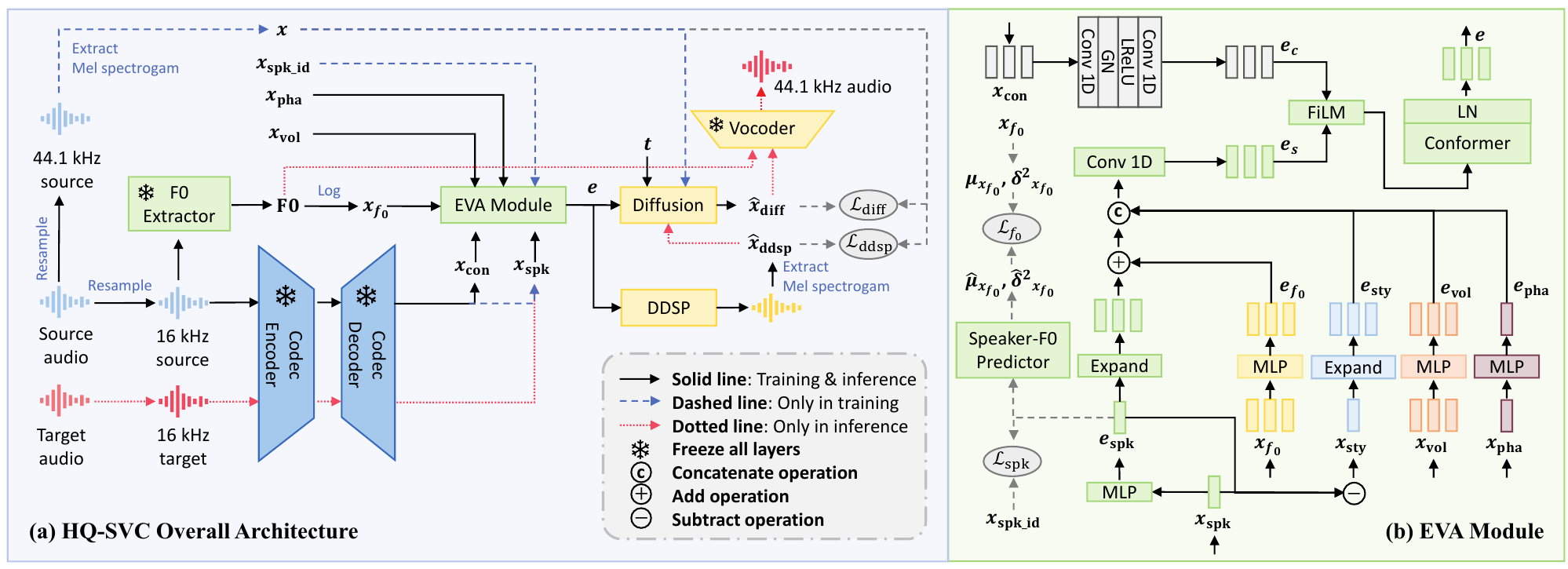}
        \caption{Overall architecture of the proposed HQ-SVC and the architecture of the proposed EVA module.}
        \label{fig:hq-svc}
\end{figure*}

As shown in Fig. \ref{fig:hq-svc} (a), HQ-SVC is built upon decoupled audio codec to achieve disentangled content features $x_\mathrm{con}$ and speaker embeddings $x_\mathrm{spk}$. EVA module is proposed to integrates other acoustic features such as pitch and energy, enhancing the capture of vocal details. Furthermore, HQ-SVC aligns with high-sampling-rate Mel spectrograms by optimizing the reconstruction loss within both DDSP and diffusion models. This enables high-quality zero-shot singing voice conversion and even voice super-resolution. In the following sections, we will provide a detailed explanation of the model architecture and the modules employed. 

\subsection{Decoupled Codec}
We utilize FACodec \cite{junaturalspeech3} as the unified disentangler, due to its high generation quality and strong disentanglement capability. Specifically, we freeze all of the layers of the encoder and decoder of the FACodec and get the $x_\mathrm{con}$ and $x_\mathrm{spk}$ as the content and speaker features respectively. The basic architecture of FACodec follows Kumar et al. \cite{kumar2024_improved_rvqgan} and employs the SnakeBeta activation function \cite{lee2023bigvgan}. The speaker extractor consists of several Conformer \cite{gulati20_interspeech_conformer} blocks. For the three factorized vector quantization (FVQ) components $Q^c$, $Q^p$, $Q^d$, we use $N_{q_c}$ = 2, $N_{q_p}$ = 1, $N_{q_d}$ = 3 quantizers respectively, with all quantizers having a codebook size of 1024.

\subsection{DDSP Model}
DDSP \cite{engel2020ddsp} is a method that integrates classical digital signal processing (DSP) techniques with deep learning. It introduces strong inductive biases while preserving the expressive power of neural networks, enabling high-fidelity audio synthesis without relying on large autoregressive models or adversarial losses. Its modular architecture allows for fine-grained control over attributes such as timbre and loudness. Specifically, the DDSP synthesizer consists of a harmonic synthesizer and a noise synthesizer, which generate periodic and aperiodic components respectively and combine them to produce the final waveform. Incorporating DDSP into the SVC task enhances control over audio attributes, thereby improving conversion performance.

\subsection{Diffusion Model}
Moreover, to improve the quality of vocal synthesis, the diffusion \cite{ho2020ddpm} model is utilized to process the output of DDSP. Integrating the diffusion model with DDSP enriches the description of sound characteristics by filling in any gaps, thereby boosting the performance of SVC. The strong performance of diffusion model in SVC tasks has been demonstrated in DiffSVC \cite{liu2021diffsvc} and CoMoSVC \cite{lu2024comosvc}.

\subsection{Enhanced Voice Adaption Module}
Singing voices are often rich in melody and energy dynamics, making it challenging to ensure high-quality generation using only content and speaker features. To address this, we propose the Enhanced Voice Adaptation (EVA) module, which introduces RMVPE \cite{wei2023rmvpe} as an F0 encoder to extract pitch features and incorporates energy features along with phase features required by the DDSP \cite{engel2020ddsp} model for multi-feature fusion. 
\subsubsection{Acoustic Feature Extraction and Mapping} As shown in Fig. \ref{fig:hq-svc} (b), since FACodec’s speaker features $x_\mathrm{spk}$ may include unrelated characteristics like style, they are mapped to speaker embeddings $e_\mathrm{spk}$ using an MLP. And the residual between the speaker features $x_\mathrm{spk}$ and speaker embeddings $e_\mathrm{spk}$ is computed to obtain residual style features $x_\mathrm{sty}$, which is expanded to yield $e_\mathrm{sty}$ that matches the sequence length dimension. Following the approach in DDSP-SVC \cite{ddsp-svc2023}, we perform the transformation on the  F0 to obtain 
\begin{equation}
x_{f_0} = \log_{e}(\frac{f_0}{700}+1),
\end{equation}
to help the model better capture the melody. And we map it to pitch embeddings $e_{f_0}$ through an MLP. Similarly, volume feature $x_\mathrm{vol}$ and phase $x_\mathrm{pha}$ feature are fed into MLPs to generate their respective embeddings $e_\mathrm{vol}$ and $e_\mathrm{pha}$. All the above MLPs consist of two linear layers, with a SiLU \cite{elfwing2018silu} activation function applied between them. The output dimension of MLPs is 256.

\subsubsection{Fusion of Style and Content Embeddings} Given the close relationship between pitch and timbre, while other features remain relatively independent, we combine the speaker and f0 embeddings into a unified representation. This combined representation is concatenated with the remaining embeddings to form a 1024-dimensional style embedding 
\begin{equation}
e_s=\mathrm{Concat}(e_\mathrm{spk}+e_{f_0},e_\mathrm{sty},e_\mathrm{vol},e_\mathrm{pha}), 
\end{equation}
which is subsequently compressed to 256 dimensions via a 1D convolution to match the content. The content features $x_\mathrm{con}$, after convolution and normalization, are fused with the aforementioned style embeddings using FiLM \cite{perez2018film}, as described by
\begin{equation}
\mathrm{FiLM}(e_c,e_s) = f_\alpha(e_s) \cdot e_c + f_\beta(e_s),\label{eq:1}
\end{equation}
where $e_c$ denotes the content embeddings, $e_s$ represents the style embeddings, and $f_\alpha(\cdot)$ and $f_\beta(\cdot)$ correspond to two independent linear layers. To achieve more effective fusion of features, the FiLM output is passed through a Conformer \cite{gulati20_interspeech_conformer} with 8-head self-attention \cite{vaswani2017attention} and layer normalization, producing $e$, which is then used as input for the DDSP and diffusion models.

In addition, to enhance the model's performance in zero-shot singing voice conversion for unseen speakers, we introduce the speaker loss $\mathcal{L}_{\mathrm{spk}}$ based on the InfoNCE loss \cite{oord2018infonce}, a widely-used contrastive learning objective. It helps the model distinguish speakers by pulling positive samples closer and pushing negative samples farther within a batch:
\begin{equation}
\mathcal{L}_{\mathrm{spk}}=-\frac{1}{N} \sum_{i=1}^N \log \left(\frac{\exp \left(\frac{e_{\mathrm{spk}, i} \cdot e_{\mathrm{spk}, j}{ }^{+}}{\tau}\right)}{\sum_{j=1}^N \exp \left(\frac{e_{\mathrm{spk}, i} \cdot {e}_{\mathrm{spk}, j}{ }^{-}}{\tau}\right)}\right),
\end{equation}
where $e_{\mathrm{spk},i}$ represents embedding of the $i$-th speaker. The positive sample $e_{\mathrm{spk}, j}{ }^{+}$ matches the same speaker ID $x_\mathrm{spk\_id}$, while negative sample $e_{\mathrm{spk}, j}{ }^{-}$ comes from different speaker IDs. $\tau$ is the temperature parameter that controls the sharpness of the similarity distribution. $N$ is the batch size.

The strong coupling between pitch range and timbre makes it essential to obtain statistical characteristics of pitch to generate singing voices that closely match the timbre of the target speaker. However, in zero-shot singing voice conversion tasks, it is typically infeasible to acquire sufficient pitch data from the target speaker to derive them. To address this, we designed a simple yet effective Speaker-F0 Predictor (SFP) $f_{\rm{sfp}}(\cdot)$ based on an MLP. It uses shared layers extract features from speaker embeddings $e_{\mathrm{spk}}$, followed by separate branches to predict mean and variance 
\begin{equation}
\hat{\mu}_{x_{f_0}}, \hat{\sigma}^2_{x_{f_0}} = f_{\mathrm{sfp}}(e_{\mathrm{spk}}).
\end{equation}
Specifically, it takes $e_{\mathrm{spk}}$ as input to predict the mean and variance of the ${x_{f_0}}$, and the loss is calculated by minimizing the L1 loss between the predicted values and the actual mean and variance of the ${x_{f_0}}$ sequence of the input. The calculation formula is
\begin{equation}
\mathcal{L}_{f_0} = \mathbb{E}[\lVert \mu_{x_{f_0}} - \hat{\mu}_{x_{f_0}} \rVert_{1} + \lVert \sigma_{x_{f_0}}^2 - \hat{\sigma}_{x_{f_0}}^2  \rVert_{1} ]. \label{eq:3}
\end{equation}
\subsection{Singing Voice Reconstruction}
We first pass the embedding $e$ from EVA module through DDSP model to generate audio. We then turn this audio into a Mel spectrogram $\hat{x}_\mathrm{ddsp}$, and minimize the Mean Squared Error (MSE) loss between $\hat{x}_\mathrm{ddsp}$ and the real Mel spectrogram $x$:
\begin{equation}
\mathcal{L}_\mathrm{ddsp} = \mathbb{E}[\lVert x - \hat{x}_\mathrm{ddsp} \rVert_2^2]
 .\label{eq:4}
\end{equation}
Additionally, we pass $e$ to the diffusion model for further optimization. We use the WaveNet \cite{oord2016wavenet} from FastSpeech2 \cite{ren2020fastspeech} as the diffusion denoiser, configured with 128-dimensional input features, 20 residual block layers, 512 convolutional output channels, and a 256-dimensional encoder hidden layer. The formula for the diffusion loss is given by:
\begin{equation}
\mathcal{L}_\mathrm{diff} = \mathbb{E}_{t,x_0,\epsilon} \left[ \lVert \epsilon - \epsilon_\theta\left( \sqrt{\bar{\alpha}_t}x_0 + \sqrt{1 - \bar{\alpha}_t}\epsilon, t, e \right) \rVert_2^2 \right],
\label{eq:5}
\end{equation}
where $\theta$ denotes the model parameters, $\epsilon$ denotes random noise, $x_0$ denotes the clean Mel spectrogram, $\bar{\alpha}_t$ is a parameter in the model, and $t$ is the time step. The expectation expression encapsulates the expectation value over all possible $t$, $x_0$, and $\epsilon$. Thus, the total loss of HQ-SVC is
\begin{equation}
\mathcal{L}_\mathrm{total}= \mathcal{L}_\mathrm{ddsp}+\mathcal{L}_\mathrm{diff} + \mathcal{L}_\mathrm{spk} + \mathcal{L}_{f_0}. \label{eq:6}
\end{equation}

During inference, DPM-Solver++ \cite{lu2022dpm} is used for sampling in the diffusion model, with chunk-wise processing adopted to reduce inference latency, 100 diffusion steps and a speed-up rate of 10 to balance quality and speed. To improve synthesis quality, HQ-SVC uses NSF-HiFiGAN \cite{openvpi2022nsf-hifigan}, an enhanced version of HiFi-GAN \cite{kong2020hifigan} with a Neural Source-Filter (NSF) \cite{wang2019nsf} module, to convert Mel spectrogram with F0 into audio.

\section{Experiments}
\label{sec:experiments}
\subsection{Datasets}
\subsubsection{Training Datasets}
In this study, we utilized the high-quality, open-source Mandarin singing datasets Opensinger \cite{huang2021opensinger} and M4Singer \cite{zhang2022m4singer} for model training. From Opensinger, 2 male and 2 female singers were randomly selected as unseen singers. These singers were excluded from both training and validation and were only used for model testing. Additionally, 1\% of the remaining singing data from M4Singer and Opensinger was randomly selected as a validation set to monitor model convergence. To avoid potential noise and performance degradation caused by excessively short audio samples, we excluded audio clips shorter than 2.1 seconds.

\subsubsection{Data Preprocessing}
To enable zero-shot singing voice conversion in low-resource scenarios, we utilized frozen pre-trained models for feature extraction, effectively reducing memory consumption during training and inference. Since the singing audio files have two different rates: 44.1 kHz and 48 kHz, we resampled them to 44.1 kHz. Then, we used a hop size of 512 and obtained 128-dimensional Mel spectrograms and energy features from the audios. Next, we downsampled the audios to 16 kHz, used RMVPE to extract pitch features, and employed the FACodec \cite{junaturalspeech3, amphion} to extract 256-dimensional content and speaker features. Furthermore, we constructed a lookup table to map all speakers across datasets to unique speaker IDs in a unified table for speaker loss calculation.

\begin{table*}[t]
  \centering
  \captionsetup{justification=centering}
  \caption{Objective and subjective evaluation results of zero-shot singing voice conversion.
           \textbf{Bold} and \underline{underline} values indicate the best and second best results, respectively.
           Confidence interval for MOS scores is 95\%.}
  \label{tab:zsvc_all}
  \begin{adjustbox}{max width=\textwidth}
    \rmfamily
    \setlength{\tabcolsep}{3pt}   
    \begin{tabular}{l|cc|ccccc|cc}
      \toprule
      \multirow{2}{*}{\textbf{Method}}
      & \multirow{2}{*}{\textbf{Train Config.}}
      & \multirow{2}{*}{\textbf{Dataset}}
      & \textbf{STOI} & \textbf{SECS} & \textbf{F0 RMSE} & \textbf{FPC} & \textbf{NISQA}
      & \textbf{NMOS} & \textbf{SMOS} \\
      & & & \textbf{(↑)} & \textbf{(↑)} & \textbf{(↓)} & \textbf{(↑)} & \textbf{(↑)}
      & \textbf{(↑)} & \textbf{(↑)} \\
      \midrule
      FACodec-SVC
      & RTX 3090 (1 h) & $<$80 h
      & 0.533 & 0.074 & 77.798 & 0.601 & 1.791
      & 2.391 ± 0.201 & 2.740 ± 0.192 \\
      SaMoye-SVC
      & A100 (7 days) & 1700 h
      & 0.724 & \textbf{0.647} & 17.418 & 0.617 & 3.528
      & 3.958 ± 0.154 & 3.569 ± 0.147 \\
      \midrule
      HQ-SVC (Ours)
      & RTX 3090 (11 h) & $<$80 h
      & \underline{0.799} & \underline{0.627} & \textbf{8.681} & \textbf{0.891} & \textbf{3.841}
      & \textbf{4.215 ± 0.124} & 3.578 ± 0.192 \\
      HQ-SVC w/o $\mathcal{L}_\mathrm{spk}$
      & RTX 3090 (11 h) & $<$80 h
      & \textbf{0.800} & 0.622 & \underline{8.828} & \underline{0.889} & 3.734
      & 4.174 ± 0.135 & \underline{3.688 ± 0.183} \\
      HQ-SVC w/o $\mathcal{L}_{f0}$
      & RTX 3090 (11 h) & $<$80 h
      & \textbf{0.800} & 0.622 & 8.855 & \underline{0.889} & \underline{3.764}
      & \underline{4.195 ± 0.138} & \textbf{3.729 ± 0.168} \\
      \bottomrule
    \end{tabular}
  \end{adjustbox}
\end{table*}

\begin{figure*}[t]
    \centering
        \includegraphics[width=\linewidth]{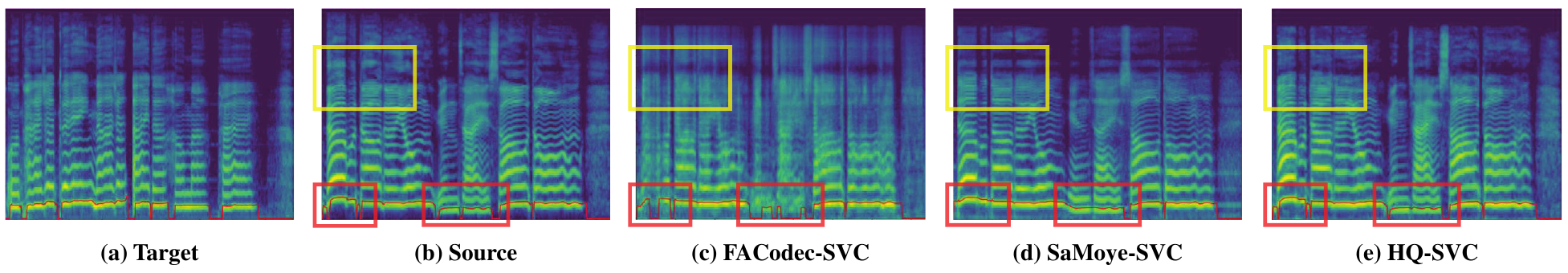}
        \caption{Spectral comparison for zero-shot singing voice conversion tasks.
The red curve represents the F0 contour, with red boxes highlighting differences in the F0 regions across the spectrograms. Yellow boxes indicate high-frequency harmonics.}
        \label{fig:svc}
\end{figure*}

\subsection{Experimental Setups}
\subsubsection{Model Training}
During training, we employed the AdamW \cite{loshchilov2017adamw} optimizer with hyperparameters \(\beta_1 = 0.9\), \(\beta_2 = 0.999\), an initial learning rate of \(1.5 \times 10^{-4}\), a batch size of \(N=64\). For the speaker loss, the temperature parameter was set to \(\tau = 0.1\). The model was trained for 250k steps on a single NVIDIA RTX 3090 GPU. 
Notably, HQ-SVC occupies less than 6 GB of GPU memory during training and can be trained within about 11 hours, making it well-suited for zero-shot singing voice conversion in low-resource scenarios.
\subsubsection{Model Evaluation on Zero-shot Singing Voice Conversion}

We conducted a comparative evaluation of the proposed HQ-SVC, against the SOTA zero-shot singing voice conversion model. Since most zero-shot SVC methods are not open-source, the VITS-based SaMoye-SVC \cite{wang2024samoye} model pretrained on 1700 hours of clean singing dataset, was used as a strong baseline for comparison, whereas HQ-SVC was trained on less than 80 hours of singing data. To further validate the effectiveness of our improvements to the FACodec, we fine-tuned FACodec. For fairness, we modified FACodec's output structure to produce Mel spectrograms, which were combined with F0 and synthesized into singing audio using NSF-HiFiGAN \cite{openvpi2022nsf-hifigan}, like HQ-SVC. In addition, all generated audios were resampled to 32 kHz for zero-shot singing voice conversion evaluation, since HQ-SVC generates audios at 44.1 kHz, while SaMoye outputs at 32 kHz. 

We assessed the models on unseen speakers using objective measures—intelligibility (STOI), timbre similarity (SECS), pitch accuracy (F0 RMSE and FPC), overall quality (NISQA) \cite{mittag2021nisqa} and subjective 5-point ratings of naturalness (NMOS) and speaker similarity (SMOS).

For MOS ratings, 12 volunteers participated in the evaluation. To evaluate zero-shot singing voice conversion methods in unseen scenarios, we randomly selected 4 audio samples from 2 male and 2 female singers in the Opensinger test set, giving us 8 source and 8 target audio samples. By performing cross-conversion between all source and target samples, we generated a total of 128 audio samples for objective evaluation. For subjective evaluation, 8 samples were randomly selected, totaling 96 per model. Fewer samples were used compared to the objective tests to maintain evaluator focus and ensure more consistent and reliable scoring.

\subsubsection{Model Evaluation on Voice Super-resolution}
HQ-SVC utilizes 16 kHz data for feature extraction during training and employs authentic 44.1 kHz Mel spectrograms for model optimization. This design allows HQ-SVC to achieve zero-shot voice super-resolution (SR) using the zero-shot singing voice conversion method while preserving the speaker's identity. To demonstrate this, we compared HQ-SVC with the current SOTA method AudioSR \cite{liu2024audiosr}. AudioSR uses a Latent Diffusion Model (LDM) trained on over 7,000 hours of sound effects, speech, and music, allowing it to upscale audio to 48 kHz. 

Since HQ-SVC uniformly resamples ground-truth audio to 44.1 kHz during training, we set its output sampling rate to 44.1 kHz for a fair comparison. We ensure fairness and demonstrate the generalization of HQ-SVC on Voice Super-resolution tasks for both singing and speech using unseen test datasets. These include Mandarin Opensinger and English NHSS-Song \cite{sharma2021nhss} (singing datasets), as well as English LibriTTS-P \cite{kawamura24_libritts_p} and NHSS-Speech (speech datasets).

For evaluation on voice super-resolution task, we randomly selected 2 male and 2 female speakers from each of the four datasets, totaling 8 male and 8 female speakers. From each speaker, we randomly chose 8 audio samples, giving us 128 samples for objective evaluation. For subjective evaluation, we randomly selected 1 audio sample per speaker, resulting in 8 singing and 8 speech samples. This yielded a total of 96 singing and 96 speech samples per model. The objective metrics include STOI, SECS, FPC, NISQA, and log-spectral distance (LSD) \cite{1976_lsd_distance_measures} metric adopted by AudioSR, while the subjective metrics include NMOS and SMOS.
\subsection{Experimental Results and Analysis}
    \subsubsection{Zero-shot Singing Voice Conversion Comparison}
The experimental results for zero-shot singing voice conversion are presented in Table \ref{tab:zsvc_all}. First, a comparison between HQ-SVC and FACodec-SVC reveals that HQ-SVC significantly outperforms FACodec-SVC across all metrics. This demonstrates the effectiveness of integrating the EVA module for multi-feature fusion, as well as leveraging DDSP and diffusion models for performance optimization. Furthermore, despite being trained in low-resource scenarios, HQ-SVC outperforms SaMoye, which was trained on large-scale datasets for an extended period. HQ-SVC achieves superior results in objective metrics such as STOI and demonstrates notable advantages in F0 RMSE, FPC, and NISQA. Additionally, it maintains an advantage in subjective metrics like NMOS and SMOS, indicating that HQ-SVC achieves higher performance in preserving melodic consistency, as well as enhancing the naturalness and similarity of singing voices.

However, we noticed that while SaMoye performs slightly worse in SMOS, it performs better in the SECS metric. This difference may be due to how SECS and human perception respond to pitch. SECS is calculated using cosine similarity from speaker embeddings extracted by the CAM++ \cite{wang23ha_interspeech_campp} speaker verification model. These embeddings implicitly contain statistical information about pitch, allowing model to effectively identify the speaker in the converted audio, even when the pitch remains unchanged. In contrast, human perception of speaker identity is more sensitive to variations in pitch. Thus, even if the speaker characteristics are altered, a consistent pitch can make it difficult for humans to identify the speaker.

The results of the ablation experiments further validate the above observations. Removing either the $\mathcal{L}_{f_0}$ or the $\mathcal{L}_\mathrm{spk}$ leads to a varying degrees of decline in SECS, F0 RMSE, NISQA, and NMOS. This indicates that these two losses contribute to generating singing voices with more accurate pitch, while ensuring that the synthesized voices remain natural and are more easily distinguishable by the speaker verification model. We also observed a slight improvement in SMOS when $\mathcal{L}_{f_0}$ or $\mathcal{L}_\mathrm{spk}$ is removed. This further supports the earlier point: adding $\mathcal{L}_{f_0}$ and $\mathcal{L}_\mathrm{spk}$ enhances pitch-related information in the speaker embeddings, making them more identifiable by CAM++ \cite{wang23ha_interspeech_campp} and better at maintaining a consistent F0. This consistency makes it difficult for human listeners to perceive changes in speaker characteristics, particularly when the target speaker has a significantly different vocal range, thus leading to a decrease in SMOS. Nevertheless, the timbre similarity of HQ-SVC remains superior to other baseline models.

To provide a more intuitive demonstration of HQ-SVC's performance in zero-shot singing voice conversion, we visualized and analyzed the spectrograms, as shown in Fig. \ref{fig:svc}. Comparing the F0 contours of the source audio and those generated by different models, we found that HQ-SVC's F0 is closest to the source, while SaMoye has overly smooth F0 contours and FACodec-SVC shows large fluctuations, both deviating from the source. These findings align with the objective experimental results, where HQ-SVC outperformed SaMoye and FACodec-SVC in F0 RMSE and FPC metrics. For high-frequency harmonics, SaMoye also shows overly smoothing, creating an unnatural sound, while FACodec-SVC adds significant noise, making the singing voice less clear. These spectrogram findings align with the objective and subjective results mentioned above.

\begin{figure}[t]
    \centering
        \includegraphics[width=\linewidth]{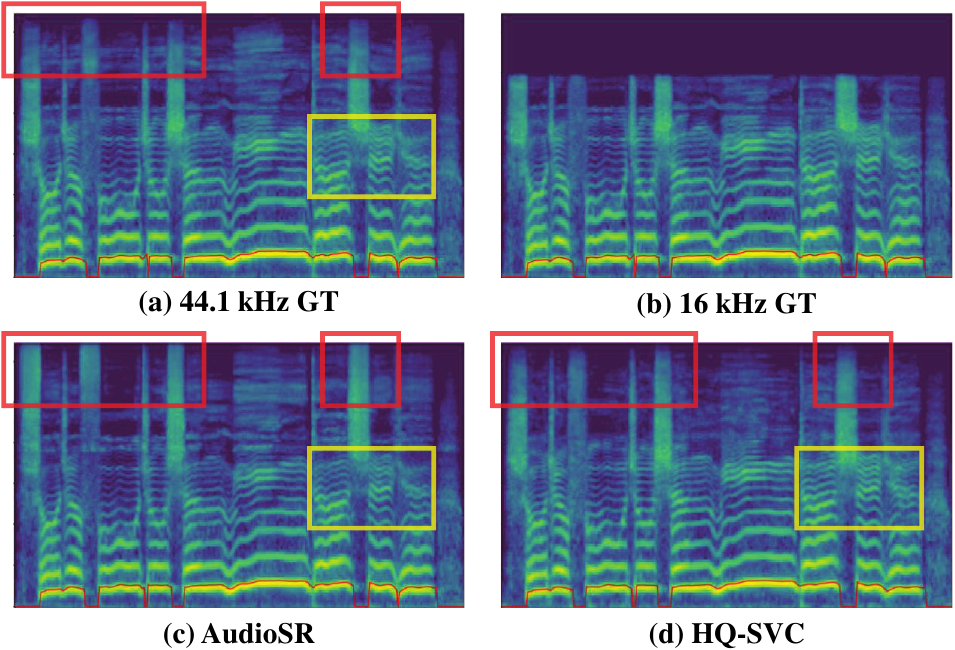}
        \caption{Spectral comparison for voice super-resolution tasks.}
        \label{fig:sr}
\end{figure}

\subsubsection{Voice Super-resolution Comparison} 

\begin{table}[t]
  \captionsetup{justification=centering}
  \caption{Objective evaluation results of voice super-resolution.}
  \label{tab:sr_obj}
  \centering
  \rmfamily

  \setlength{\tabcolsep}{1pt}
  \resizebox{0.9\columnwidth}{!}{
  \begin{tabular}{@{\extracolsep{\fill}}l|ccccc}
    \toprule
    \textbf{Method} & \textbf{STOI(↑)} & \textbf{SECS(↑)} & \textbf{FPC(↑)} &\textbf{LSD(↓)} & \textbf{NISQA(↑)}\\
    \midrule
    Ground Truth  & -  & - & - & - & 4.234 \\
    Downsampled & - & - & - & - & 4.181\\
    \midrule
    AudioSR  & \textbf{0.986} & 0.759 & \textbf{0.998} & 2.087 & 4.094\\
    \midrule
    HQ-SVC (Ours) & 0.841 & \textbf{0.766} & 0.868 & \textbf{1.842} & \textbf{4.193 }\\
    \bottomrule
  \end{tabular}
  }
\end{table}

\begin{table}[t]
  \captionsetup{justification=centering}
  \caption{Subjective evaluation results of voice super-resolution with the confidence interval 95\%.}
  \label{tab:sr_sub}
  \centering
  \rmfamily

  \setlength{\tabcolsep}{3pt}
  \resizebox{0.9\columnwidth}{!}{ 
  \begin{tabular}{@{\extracolsep{\fill}}l|c|cc}
    \toprule
    \textbf{Method} & \textbf{Dataset} & \textbf{NMOS(↑)} & \textbf{SMOS(↑)} \\
    \midrule
    Ground Truth  & - & 4.667 ± 0.089 & - \\
    Downsampled & - & 4.083 ± 0.100 & - \\
    \midrule
    AudioSR  & 7000 h & 4.188 ± 0.103 & 4.235 ± 0.096 \\
    \midrule
    HQ-SVC (Ours) & $<$80 h & \textbf{4.332 ± 0.088} & \textbf{4.479 ± 0.087}  \\
    \bottomrule
  \end{tabular}
  }
\end{table}

\begin{table}[ht]
  \centering
  \caption{Performance comparisons in different sampling methods at different acceleration in singing voice conversion tasks.}
  \label{tab:vc_acc}
  \begin{adjustbox}{max width=\linewidth} 
    \begin{tabular}{lc|cccccc}
      \toprule
      \textbf{Method} & \textbf{Speed} & \textbf{RTF} & \textbf{STOI} & \textbf{SECS} & \textbf{F0 RMSE} & \textbf{FPC} & \textbf{NISQA} \\
                      &                & {(↓)}        & {(↑)}         & {(↑)}         & {(↓)}            & {(↑)}        & {(↑)}         \\
      \midrule
      \multirow{4}{*}{Dpm-solver++} 
        & 1×  & 0.295  & 0.801  & 0.620  & \underline{8.578}  & 0.888  & 3.803 \\
        & 5×  & 0.090  & 0.799  & \underline{0.624}  & \textbf{8.575}  & 0.888  & \underline{3.825} \\
        & 10× & 0.065  & 0.799  & \textbf{0.627}  & 8.681  & \textbf{0.891}  & \textbf{3.841} \\
        & 20× & 0.054  & \textbf{0.808}  & 0.570  & 8.695  & 0.889  & 3.684 \\
      \midrule
      \multirow{4}{*}{UniPC}
        & 1×  & 0.275  & 0.801  & 0.619  & 8.611  & 0.888  & 3.798 \\
        & 5×  & 0.085  & 0.802  & 0.612  & \underline{8.578}  & 0.889  & 3.775 \\
        & 10× & 0.061  & \underline{0.804}  & 0.598  & 8.699  & \underline{0.890}  & 3.737 \\
        & 20× & \underline{0.048}  & \textbf{0.808}  & 0.570  & 8.767  & 0.889  & 3.686 \\
      \midrule
      \multirow{4}{*}{DDIM}
        & 1×  & 0.203  & 0.801 & 0.617  & 8.652  & 0.888  & 3.789 \\
        & 5×  & 0.067  & 0.801  & 0.613  & 8.745  & 0.889  & 3.773 \\
        & 10× & 0.060  & 0.801  & 0.612  & 8.754  & 0.889  & 3.763 \\
        & 20× & \textbf{0.043}  & 0.799  & 0.618  & 8.769  & 0.889  & 3.782 \\
      \bottomrule
    \end{tabular}
  \end{adjustbox}
\end{table}

\begin{table}[ht]
  \setlength{\tabcolsep}{4pt} 
  \captionsetup{justification=centering}
  \caption{Ablation studies in singing voice conversion.}
  \label{tab:ablation}
  
  \centering
  \rmfamily
  \resizebox{1\columnwidth}{!}{ 
    \begin{tabular}{l|ccccc}
      \toprule
      \textbf{Method} & \textbf{STOI(↑)} & \textbf{SECS(↑)} & \textbf{F0 RMSE(↓)} & \textbf{FPC(↑)} & \textbf{NISQA(↑)} \\
      \midrule    
      HQ-SVC & 0.799 & 0.627 & 8.681 & 0.891 & 3.841 \\
      w/o Diffusion & 0.820 & 0.420 & 8.547 & 0.888 & 3.175 \\
      w/o DDSP & 0.847 & 0.608 & 8.224 & 0.901 & 3.960 \\
      \midrule
      HQ-SVC-SE & 0.738 & 0.668 & 9.803 & 0.873 & 3.880 \\
      w/o $\mathcal{L}_\mathrm{spk}$ & 0.735 & 0.660 & 9.747 & 0.873 & 3.780 \\
      w/o $\mathcal{L}_{f0}$ & 0.739 & 0.663 & 9.814 & 0.871 & 3.790 \\
      \bottomrule
    \end{tabular}
  }
\end{table}

The objective experimental results for voice super-resolution are shown in Table \ref{tab:sr_obj}. HQ-SVC shows clear advantages in metrics like LSD and NISQA, producing higher-quality voices closer to the ground truth. These results highlight HQ-SVC's ability to generate high-fidelity audio and effectively reconstruct details in voice super-resolution tasks, even in low-resource situations. However, HQ-SVC performs poorly in STOI and FPC compared to AudioSR, indicating that AudioSR produces audio with better intelligibility in terms of F0 and content. This difference is mainly due to the distinct training methods of the models. AudioSR, a specialized audio super-resolution model, is trained on downsampled and high-sample-rate Mel spectrogram pairs, helping it better preserve low-frequency details like pronunciation and F0. In contrast, HQ-SVC, as a zero-shot singing voice conversion model, focuses on controllability and generates high-sample-rate audio by decoupling features and directly reconstructing them into Mel spectrograms. While this approach offers more flexibility and control, it introduces minor and acceptable errors in pronunciation and pitch.

To evaluate the subjective quality of audio generated by HQ-SVC, we conducted an MOS evaluation, as presented in Table \ref{tab:sr_sub}. The results show that HQ-SVC outperforms AudioSR in both NMOS and SMOS, confirming its ability to produce high-quality and natural voice super-resolution in low-resource scenarios.

The spectral comparison in Fig. \ref{fig:sr} shows that HQ-SVC upscales low-sample-rate speech to 44.1 kHz with clearer mid-to-high frequency harmonics (yellow boxes) and less high-frequency noise (red boxes) than AudioSR, generating more natural voices. This is mainly due to the HQ-SVC, which does not simply rely on increasing the signal-to-noise ratio to reconstruct the audio. Instead, it leverages multi-feature fusion, making the audio sound more like natural voice. This aligns with our previous objective and subjective results.

\subsubsection{Sampler and Acceleration Rate Comparison in HQ-SVC}

Table \ref{tab:vc_acc} presents a comparison of zero-shot singing voice conversion performance under various acceleration rates (1×, 2×, 10×, 20×) using three diffusion-based samplers: Dpm-solver++ \cite{lu2022dpm}, UniPC \cite{zhao2023unipc}, and DDIM \cite{song2020ddim}. 

We compare diffusion samplers and acceleration rates using RTF, STOI, SECS, F0 RMSE, FPC, and NISQA. In this study, which focuses on comparing the performance of different samplers and acceleration rates, we define RTF as the ratio of the time taken to convert the embeddings into audio to the duration of the generated audio. The results show that Dpm-solver++ at 10× gives the best speed-quality trade-off.

\subsubsection{Ablation Studies}
We did more ablation studies to check how well each part works and to better understand the results. As shown in Table \ref{tab:ablation}, taking out the diffusion model made speech a bit clearer but much worse in terms of matching the original voice (timbre) and natural sound. This shows the diffusion model is important for keeping the speaker’s voice and making converted speech sound natural. On the other hand, removing the DDSP module improved some scores but also made timbre matching worse. This might look like it’s better for singing, but because the diffusion model uses the original audio’s Mel spectrogram, the output stays close to the original. So even with higher scores, it doesn’t actually convert voices well.

We also compared HQ-SVC with a version using separate encoders (HQ-SVC-SE). In HQ-SVC-SE, we replaced FACodec with two separate encoders: CAM++ (for speakers) and ContentVec (for content). ContentVec is based on HuBERT and is designed to remove timbre info. The comparison showed HQ-SVC’s single encoder is better at combining different types of information and controlling rhythm. But HQ-SVC still isn’t as good as HQ-SVC-SE at matching timbre, which means we need to get better at separating out the speaker’s identity. Future work could use methods like changing pitch or speaker characteristics to improve this. Also, removing either the speaker loss or pitch loss hurt how well the model handled pitch (F0) and timbre matching, showing these are important for good voice conversion.

In short, HQ-SVC balances performance across different measures. It might not be the best in every single area, but its complete design, reliability, and ability to work in different situations make it the best choice as the core of our system.

\section{Conclusions}

\label{sec:conclusions}
This paper introduces HQ-SVC, a zero-shot singing voice conversion method for low-resource scenarios. It uses a decoupled codec to disentangle content and speaker features and integrates additional acoustic features with the proposed Enhanced Voice Adaption module. Both objective and subjective experiments show that HQ-SVC achieves high naturalness and similarity in zero-shot singing voice conversion, and even performs well in zero-shot voice super-resolution, despite limited GPU resources and a small training dataset. Future work will focus on singing voice style conversion.

\section{Acknowledgments}
\label{sec:acknowledgments}
This work is supported by the National Key R\&D Program of China under Grant No.2024YFB2808802.

\bibliography{main}
\end{document}